\documentclass[preprint,12pt,authoryear]{elsarticle}

\usepackage{amssymb}
\usepackage{amsmath}
\usepackage{makecell}
\usepackage{pdflscape}
\usepackage{graphicx}
\usepackage{booktabs} 
\usepackage[table]{xcolor}
\usepackage{microtype}
\usepackage{subcaption}

\journal{Finance Research Letter}

\begin{document}

\begin{frontmatter}



\title{Unveiling Hedge Funds: Topic Modeling and Sentiment Correlation with Fund Performance} 


\author{Chang Liu} 

\affiliation{organization={Department of Information Engineering and Computer Science, University of Trento},
            addressline={Via Sommarive, 9}, 
            city={Povo},
            postcode={38123}, 
            state={Trient},
            country={Italy}}

\begin{abstract}
		The hedge fund industry presents significant challenges for investors due to its opacity and limited disclosure requirements. This pioneering study introduces two major innovations in financial text analysis. First, we apply topic modeling to hedge fund documents-an unexplored domain for automated text analysis-using a unique dataset of over 35,000 documents from 1,125 hedge fund managers. We compared three state-of-the-art methods: Latent Dirichlet Allocation (LDA), Top2Vec, and BERTopic. Our findings reveal that LDA with 20 topics produces the most interpretable results for human users and demonstrates higher robustness in topic assignments when the number of topics varies, while Top2Vec shows superior classification performance. Second, we establish a novel quantitative framework linking document sentiment to fund performance, transforming qualitative information traditionally requiring expert interpretation into systematic investment signals. In sentiment analysis, contrary to expectations, the general-purpose DistilBERT outperforms the finance-specific FinBERT in generating sentiment scores, demonstrating superior adaptability to diverse linguistic patterns found in hedge fund documents that extend beyond specialized financial news text. Furthermore, sentiment scores derived using DistilBERT in combination with Top2Vec show stronger correlations with subsequent fund performance compared to other model combinations. These results demonstrate that automated topic modeling and sentiment analysis can effectively process hedge fund documents, providing investors with new data-driven decision support tools.
\end{abstract}


\begin{highlights}
\item First application of topic modeling techniques to hedge fund documents, analyzing over 35,000 documents from 1,125 managers.
\item Three topic modeling techniques—LDA, Top2Vec, and BERTopic—are proposed to identify topics in the hedge fund industry. LDA with 20 topics produces the most interpretable results, while Top2Vec shows superior classification performance.
\item Sentiment analysis is performed using DistilBERT and FinBERT to evaluate fund performance correlations. Documents' sentiment score significantly correlate with future fund performance, establishing a novel quantitative indicator.
\end{highlights}


\end{frontmatter}


\section{Introduction}\label{introduction}

    The hedge fund industry is one of the most opaque and least transparent sectors within the financial industry. Due to limited regulatory disclosure requirements and proprietary investment strategies \citep{RePEc:bla:jfinan:v:63:y:2008:i:6:p:2785-2815}, investors face considerable challenges in evaluating hedge funds. To make informed investment decisions, they must rely on fund-related documents. Traditionally, investors assess hedge fund documents based on several key factors, including historical performance, investment strategy, team composition, organizational structure, and fee structures \citep{DUNN2006363}. These factors help benchmark funds against industry standards and facilitate peer-group comparisons. However, manually reviewing and consolidating such information is not only prone to subjectivity and human bias but also highly time-consuming, particularly given the vast number of global hedge funds.

    Natural Language Processing (NLP) has emerged as a powerful tool for analyzing large-scale textual data \citep{LAURIOLA2022443}, with Topic Modeling (TM) standing out as a particularly effective technique \citep{VAYANSKY2020101582}. TM enables the automatic identification of themes and latent topics within extensive document collections, offering a scalable and unbiased approach to text mining. Unlike supervised learning methods, which require pre-labeled training data, TM operates unsupervised, making it well-suited for uncovering hidden structures in unstructured financial documents. Despite its potential, TM remains underutilized in the hedge fund industry, where manual document review is still the dominant approach. A key barrier to adoption is the selection of appropriate TM techniques. 

    Topic modeling in hedge fund document analysis offers several compelling advantages. First, it enables efficient processing of vast document collections that would be impractical to review manually. Second, it provides objective, consistent analysis that minimizes human bias in document interpretation. Third, it can potentially uncover latent relationships between document content and fund performance that might not be apparent through traditional analysis. Finally, automated topic extraction creates a foundation for downstream applications such as document classification, sentiment analysis, and information retrieval systems tailored to investor needs.


    To the best of our knowledge, no prior research has applied TM techniques to hedge fund documents. Additionally, our study is the first to investigate the correlation between funds' future performance and the sentiment scores of topics identified in the hedge fund communications. However, several related studies have explored the application of text analysis in financial contexts.
    
    This study was inspired by \citep{RePEc:usg:econwp:2023:07}, which analyzed the Management Discussion and Analysis (MD\&A) sections of 10-K filings from 1994 to 2018. The authors employed a word clustering approach based on anchor words to identify topics, extract measures of topic prevalence, and assess sentiment. Their methodology demonstrated the value of combining topic extraction with sentiment analysis in financial documents, though in the context of public company filings rather than hedge fund materials.
    
    Expanding the scope of financial text analysis, we also examined \citep{electronics12122605}, which compared LDA, Top2Vec, and BERTopic in a controlled scenario analyzing the impact of financial news on markets.  Their experimental results demonstrated that BERTopic outperformed the other models in this context, providing a methodological benchmark for our own comparative analysis.

    Other significant work in financial text analysis includes studies applying NLP techniques to earnings calls \citep{RePEc:bla:joares:v:50:y:2012:i:2:p:495-540}, \citep{eaningcall_cite2}, financial news \citep{doi:10.3233/IDT-230478}, and social media sentiment for market prediction \citep{7955659}, \citep{BOLLEN20111}. However, the unique characteristics of hedge fund documents—including their specialized terminology, heterogeneous structure, and significant regulatory content—create distinct challenges that needs specialized investigation.


    This paper presents two significant innovations to address these challenges. First, we pioneer the application of TM to hedge fund documents - a previously unexplored domain for automated text analysis techniques. To effectively demonstrate this novel application, our study compares three state-of-the-art TM methods: Latent Dirichlet Allocation (LDA) \citep{NIPS2001_296472c9}, Top2Vec \citep{angelov2020top2vecdistributedrepresentationstopics}, and BERTopic \citep{grootendorst2022bertopic}. While LDA relies on a probabilistic generative model, Top2Vec and BERTopic leverage word embeddings to capture contextual relationships. However, as these embeddings are typically trained on general text corpora rather than domain-specific financial data, their effectiveness in the hedge fund industry remains uncertain. We evaluate these models on a unique dataset of over 35,000 documents from 1,125 hedge fund managers, assessing their effectiveness in extracting meaningful topics that align with categories relevant to investment decision-making. Our analysis includes both quantitative metrics (i.e. coherence scores, classification accuracy) and qualitative evaluation (i.e. topic interpretability, applicability to investment workflows).

    The second major innovation is establishing a novel link between document sentiment and hedge fund performance. By analyzing investment team commentaries, we compute sentiment scores at the topic level to create quantifiable metrics that could streamline the evaluation process. While relationships between market commentaries and subsequent fund performance may seem intuitive, our innovation lies in developing automated sentiment indicators that can be systematically integrated into investment workflows. This approach transforms qualitative information traditionally requiring expert interpretation into quantitative signals that can enhance fund evaluation models, potentially offering predictive power for future fund performance and providing investors with new data-driven decision support tools.


    
    The rest of the paper is organized as follows. Section \ref{topic_modelling} applies LDA, Top2Vec, and BERTopic to the hedge fund documents, analyzing the resulting topic lists through a combination of manual annotation and Large Language Model (LLM)-assisted labeling. Section \ref{senti_analysis} calculates sentiment scores at topic and fund level, investigating their correlation with subsequent fund performance. Finally, Section \ref{outlook} concludes the study and discusses directions for future research.

    \section{Topic Models on Hedge Fund Documents}
    \label{topic_modelling}
    In this paper, we constructed a unique dataset comprising 35,225 hedge fund documents from 1,125 hedge fund managers from June 30, 2017, to January 31, 2025. A detailed description of the dataset and the data cleaning process, which included text extraction, normalization, and chunking, is provided in \ref{appendix_data_preprocessing}. After preprocessing, the dataset consists of 516,152 text chunks.

    To identify latent topics in hedge fund documents and evaluate which topic modeling approach best captures the nuanced information relevant to investors, we systematically compared three state-of-the-art topic modeling techniques. We implemented LDA, Top2Vec, and BERTopic using the Python libraries sci-kit-learn, top2vec, and BERTopic, respectively. A comprehensive summary of their implementation details, including the required libraries and parameter configurations, is provided in Table \ref{tab:all_models}. We applied LDA with two different configurations, setting the number of predefined topics to 20 and 100. Unlike LDA, Top2Vec and BERTopic do not require a predefined number of topics. 
    
    Table \ref{tab:topics_comparison} presents a comparative analysis of the topics generated by the models in this study. We report the total number of topics identified by each model, noting that these values are predetermined for LDA. Additionally, we include the number of documents assigned to each topic. LDA and BERTopic provide an outlier class for documents that cannot be classified into any specific topic, whereas Top2Vec does not incorporate this feature.

    LDA 20 model generates topics with a larger number of associated documents per topic while generating fewer outliers than LDA 100. In contrast, Top2Vec and BERTopic yield an extremely large number of topics, many of which contain only a small number of classified documents. This fragmentation complicates direct interpretation. To address this, we applied hierarchical topic reduction, constraining the number of topics in Top2Vec and BERTopic to align with the configurations used in LDA. The statistics for the topic lists generated by these models are presented in the columns Top2Vec 20, Top2Vec 100, BERTopic 20, and BERTopic 100 of Table \ref{tab:topics_comparison}. The results indicate that LDA and Top2Vec effectively categorize the majority of documents into well-sized topics, whereas BERTopic assigns nearly all documents to either the outlier class or the first topic class. Based on these findings, we focus on LDA 20, LDA 100, Top2Vec 20, and Top2Vec 100 for further validation.

        \begin{landscape}
    \begin{table}
    \scriptsize
        \centering
        \begin{tabular}{lcccccc}
        \hline
             & LDA 20 & LDA 100 & Top2Vec & Top2Vec n-gram & BerTopic & BerTopic n-gram\\ 
        \hline
            No. of topics & 20 & 100 & undefined & undefined & undefined & undefined\\

            n-gram & 1 & 1 & 1 & 1-3 & 1 & 1-3 \\
            
            max iterations & 50 & 50 & undefined & undefined & undefined & undefined\\
            
            min topic size & undefined & undefined & undefined  & undefined & 10 & 10\\
                        
            dimensionality reduction & undefined & undefined & UMAP & UMAP & UMAP & UMAP\\
            
            clustering & undefined & undefined & HDBSCAN & HDBSCAN & HDBSCAN & HDBSCAN\\
            
            topic representation & default  & default &  centroid proximity &  centroid proximity &  c-TF-IDF, MMR &  c-TF-IDF, MMR \\
            \hline
        \end{tabular}
        \caption{Technical implementation details of TM approaches applied to hedge fund documents. This table outlines the specific configurations and parameters used for LDA, Top2Vec and BERTopic models in this study. "No. of topics" indicates whether the number of topics was preset or algorithmically determined (undefined). "n-gram" refers to the word sequence length considered in the analysis, with single words (1) or word sequences of 1-3 words (1-3). "Max iterations" shows the maximum number of iterations allowed for the LDA algorithm to reach convergence. "Min topic size" specifies the minimum number of paragraph text required to form a topic in BERTopic. "Dimensionality reduction" indicates the technique used to reduce high-dimensional vector spaces, with Uniform Manifold Approximation and Projection (UMAP) \citep{mcinnes2020umapuniformmanifoldapproximation} employed in Top2Vec and BERTopic for more effective visualization and clustering. "Clustering" denotes the algorithm used for grouping similar documents, with Hierarchical Density-Based Spatial Clustering of Applications with Noise (HDBSCAN) \citep{10.1007/978-3-642-37456-2_14} used in Top2Vec and BERTopic. "Topic representation" describes the method used to identify representative words for each topic: LDA uses a probabilistic approach (default), Top2Vec employs embedding centroid proximity, and BERTopic utilizes class-based Term Frequency-Inverse Document Frequency (c-TF-IDF) \citep{10.5555/106765.106782} and Maximal Marginal Relevance (MMR) to select diverse yet representative terms.}
        \label{tab:all_models}
    \end{table}
    \end{landscape}

    \begin{landscape}
    \begin{table}
    \scriptsize
        \centering
        \begin{tabular}{lcccccccccc}
        \hline
             & LDA 20 & LDA 100 & Top2Vec  & Top2Vec ngram & Top2Vec 20 & Top2Vec 100 & BerTopic  & BerTopic ngram & BerTopic 20 & BerTopic 100\\
        \hline
            No. of topics & 20 & 100 & 8,462 & 8,489 & 20 & 100 & 13,942 & 13,969 & 20 & 100\\
            outliers & 134,677 & 247,277 & undefined & undefined &  undefined & undefined & 71,172 & 71,779 &  72,775 & 72,775\\
            max \#n docs & 54,879 & 12,464 & 4,758 & 4,857 & 41,364 & 13,408 & 716 & 665 & 438,236  & 406,487\\
            median \#n docs & 16,557 & 1,918 & 38 & 38 & 22,951 & 4,776 & 24 & 24 & 73  & 96\\
            mean \#n docs & 19,074 & 2,689 & 61 & 61 & 25,808 & 5162  &  32 & 32 & 23,336 & 4,479\\
            min  \#n docs & 638 & 140 & 1 & 1 &  16,927 & 2,790 &  10 & 10 & 14  & 11\\
            \hline
        \end{tabular}
        \caption{Comparison of topic distribution statistics across different TM approaches applied to hedge fund documents. "No. of topics" indicates the total number of distinct topics identified by each model, with constrained versions (20 and 100) and unconstrained ones. "Outliers" represents the number of paragraph text that could not be assigned to any topic by the respective algorithms. "Max \#n docs" shows the size of the largest topic in each model (maximum number of paragraph text assigned to a single topic), while "Min \#n docs" shows the smalles topic size. "Median \#n docs" and "Mean \#n docs" show central tendency for topic sizes. LDA and constrained Top2Vec models (LDA 20, LDA 100, Top2Vec 20, and Top2Vec 100) produce well-balanced topic sizes suitable for human interpretation, while unconstrained models generate numerous small topics. BERTopic 20 and 100 exhibit extreme variation in topic sizes despite the topic number constraint. The substantial differences in document distribution patterns highlight the importance of model selection when applying topic modeling to hedge fund documents. Ngram variants incorporate multi-word phrases in topic extraction; details can be found in \ref{appendix_ngram}.}
        \label{tab:topics_comparison}
    \end{table}
    \end{landscape}


    \section{Validation and Interpretation of Model outputs}
    To validate the LDA and Top2Vec model outputs, we conducted a rigorous evaluation process combining automated and manual assessment. First, we examined high-frequency words and representative text samples from each generated topic. For systematic annotation, we randomly sampled text examples from each topic and employed ChatGPT to classify them into seven predefined categories: "Disclosure", "Fund Terms", "Investment Team", "Market Update", "Performance Commentary", "Strategy Overview", and "Other". We then manually reviewed these classifications to ensure accuracy.


    All topic modeling results are presented in \ref{appendix_topic_models_tables}, which contains a comprehensive set of tables documenting both the topic content and classification performance of our models. For the LDA 20 model, we provide the complete list of extracted topics with their annotations (Table \ref{tab:topics_lda_20}) and the distribution of text samples across classification categories as determined through both ChatGPT and human evaluation (Table \ref{tab:metrics_lda_20}). Similarly, for comparative analysis, the appendix includes parallel tables for LDA 100 (Tables \ref{tab:topics_lda_100} and \ref{tab:metrics_lda_100}), Top2Vec 20 (Tables \ref{tab:topics_top2vec} and \ref{tab:metrics_t2v_20}), and Top2Vec 100 (Tables \ref{tab:topics_top2vec_100} and \ref{tab:metrics_t2v_100}). This organized presentation allows readers to directly compare the performance and characteristics of each model configuration.

    As expected, many topics are categorized under "Disclosure". This predominance of disclosure-related topics stems from two primary factors. First, hedge fund documents typically contain substantial amounts of legal disclaimer and regulatory statement text. Second, these disclosure texts exhibit significant variation in structure, content, and terminology across different asset managers and document types, causing the topic modeling algorithms to recognize them as distinct topics. 
    
    Despite this disclosure-heavy distribution, LDA 20 and LDA 100 effectively identify key financial themes such as performance commentary, market updates, strategy overviews, and investment team descriptions. Similarly, the Top2Vec 20 model captures topics related to market updates, strategy updates, and performance commentary; however, it does not identify the investment team theme. While the Top2Vec 100 model successfully identifies the investment team theme, it does not capture performance commentary within its 20 most significant topics.

    Topic annotation was primarily based on sample text from each topic rather than high-frequency words. However, topics generated by LDA exhibited less overlap among their top words, making these words more intuitive indicators of the underlying topic meaning.

    To evaluate the ability to distinguish between informative and non-informative texts, we define a topic as a positive prediction if less than 50\% of its texts are classified as disclosure, and as a negative prediction otherwise. Precision, recall, and F1 scores are computed and reported in Table \ref{tab:cohhrence_scores}. LDA 20 achieves the highest precision but a lower F1 score due to reduced recall, while Top2Vec 20 attains a comparable precision with significantly higher recall.


    To further evaluate the quality of the topics, we computed coherence scores. As noted in \citep{ABDELRAZEK2023102131}, coherence scores are the most appropriate evaluation metric when topic model outputs are used for human interpretation. Table \ref{tab:cohhrence_scores} reports the coherence scores \( C_\texttt{V} \) proposed by \citep{10.1145/2684822.2685324} and \( C_{\texttt{UMass}} \) proposed by \citep{mimno-etal-2011-optimizing}. Scores are calculated with the top 10 words from each topic. The results indicate that LDA significantly outperforms the Top2Vec, with the LDA 20 achieving the highest coherence scores.
    
\begin{figure}[ht]
        \centering
        \begin{subfigure}{0.45\textwidth}
            \centering
        \includegraphics[width=1\linewidth]{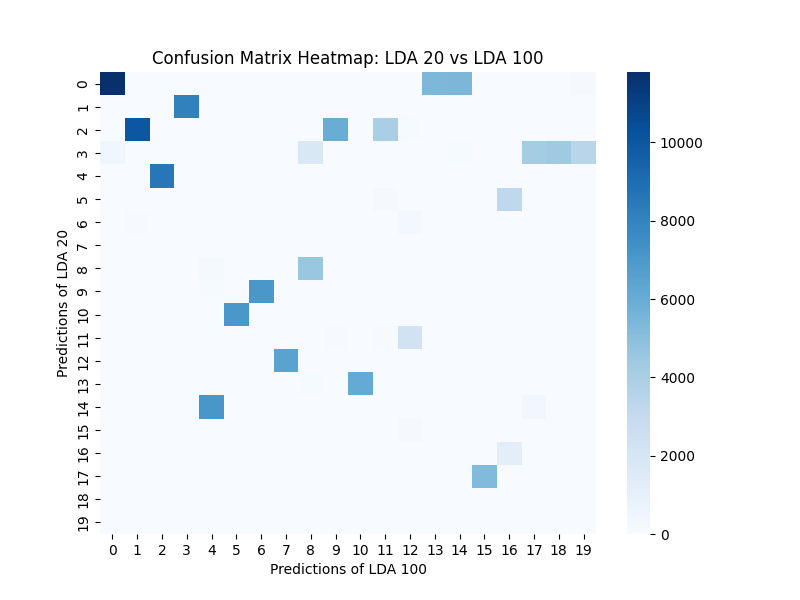}
        \caption{Confusion matrix comparing LDA 20 and LDA 100. The matrix contains few nonzero elements and minimal row or column overlap, indicating greater consistency and stability in topic assignments.}
        \label{fig:comparison_heatmap_lda_20_100}
        \end{subfigure}
        \hfill
        \begin{subfigure}{0.45\textwidth}
            \centering
        \includegraphics[width=1\linewidth]{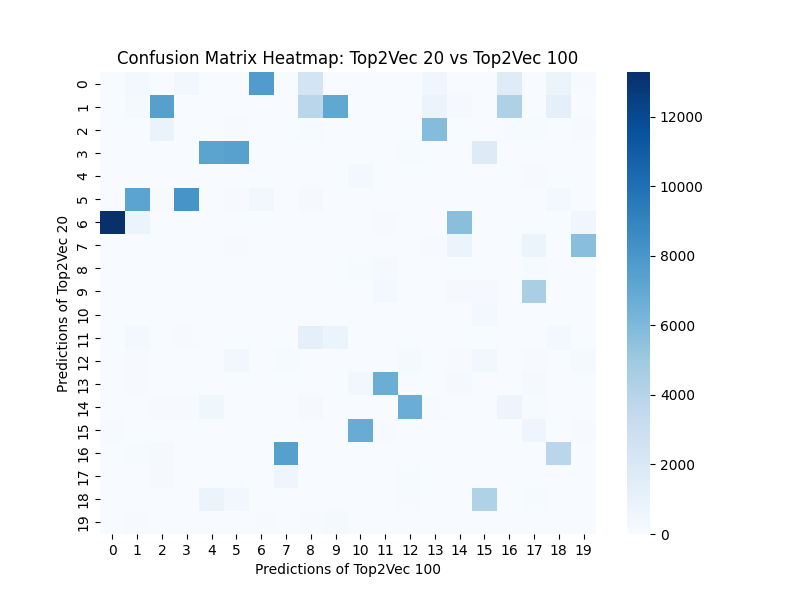}
        \caption{Confusion matrix comparing Top2Vec 20 and Top2Vec 100. The matrix contains more nonzero elements and greater row and column overlap, suggesting lower consistency and stability in topic assignments.}
        \label{fig:comparison_heatmap_top2vec_20_100}
        \end{subfigure}
        
    \caption{Confusion matrices comparing topic stability in LDA and Top2Vec models when increasing the number of topics from 20 to 100. LDA exhibits greater topic assignment stability.}
        \label{fig:comparison_heatmaps}
    \end{figure}

    To assess the robustness of our TM approaches, we visualize and compare the confusion matrices for LDA and Top2Vec. These matrices illustrate how topic assignments shift when the number of topics varies within each modeling method. Figure \ref{fig:comparison_heatmap_lda_20_100} presents the confusion matrix comparing LDA 20 and LDA 100, focusing on the first 20 topics. Similarly, Figure \ref{fig:comparison_heatmap_top2vec_20_100} provides the confusion matrix for Top2Vec 20 and Top2Vec 100. The confusion matrix for LDA contains fewer nonzero elements compared to that of Top2Vec, suggesting that LDA exhibits greater consistency in topic assignment when the number of topics varies.

    \begin{table}
        \centering
        \begin{tabular}{ccccccccccc}
        \hline
            metrics score & LDA 20 & LDA 100& Top2Vec 20 &Top2Vec 100 \\
        \hline
                    Precision & 0.9237 & 0.8855 & 0.9231 &  0.9124 \\
            Recall & 0.7331 & 0.8147 &  0.8336 & 0.8146 \\
            F1 score &  0.8174 & 0.8486 &  0.8761  &  0.8607 \\
            $C_\texttt{V}$ & 0.5442 & 0.5363 & 0.3315 & 0.3588 \\
            $U_{\texttt{UMass}}$ & -2.1801 & -2.8640 & -7.3973 & -7.8036 \\

            \hline
        \end{tabular}
        \caption{Comparative performance metrics for TM methods applied to hedge fund documents. The table compares LDA and Top2Vec with different numbers of topics (20 and 100). Precision measures how well the model correctly identifies relevant documents, while Recall indicates the proportion of relevant documents successfully retrieved. F1 score represents the harmonic mean of Precision and Recall. $C_\texttt{V}$ (Coherence score based on normalized pointwise mutual information and cosine similarity) and $U_{\texttt{UMass}}$ (Coherence score based on document co-occurrence) evaluate topic quality and interpretability, with higher $C_\texttt{V}$ and less negative $U_{\texttt{UMass}}$ values indicating better topic coherence. LDA significantly outperforms Top2Vec regarding coherence scores, with LDA 20 achieving the highest coherence, suggesting it produces the most interpretable topics for humans. However, Top2Vec 20 demonstrates superior classification performance with higher F1 scores (0.8761) compared to LDA 20 (0.8174), indicating better overall accuracy in document classification tasks.}
        \label{tab:cohhrence_scores}
    \end{table}



    \section{Sentiment Analysis and Fund Performance}
    \label{senti_analysis}
    To examine the relationship between sentiment and future fund performance, we applied sentiment analysis at the document level using two different pre-trained models from the Hugging Face transformers library. The first model, DistilBERT \citep{sanh2020distilbertdistilledversionbert}, is a general-purpose sentiment analysis model trained on broad sentiment classification tasks. The second model, FinBERT \citep{DBLP:journals/corr/abs-1908-10063}, is specifically designed for financial text analysis, making it better suited for capturing nuanced financial sentiment.

    We compute a sentiment score using both models for each document. These sentiment scores are aggregated at the topic level and averaged for each fund and month. Finally, we calculate the correlation between the averaged sentiment scores and the subsequent month's fund performance, providing insight into how sentiment within different topics relates to financial outcomes.

\begin{figure}[ht]
        \centering
        \begin{subfigure}{0.45\textwidth}
            \centering
            \includegraphics[width=\linewidth]{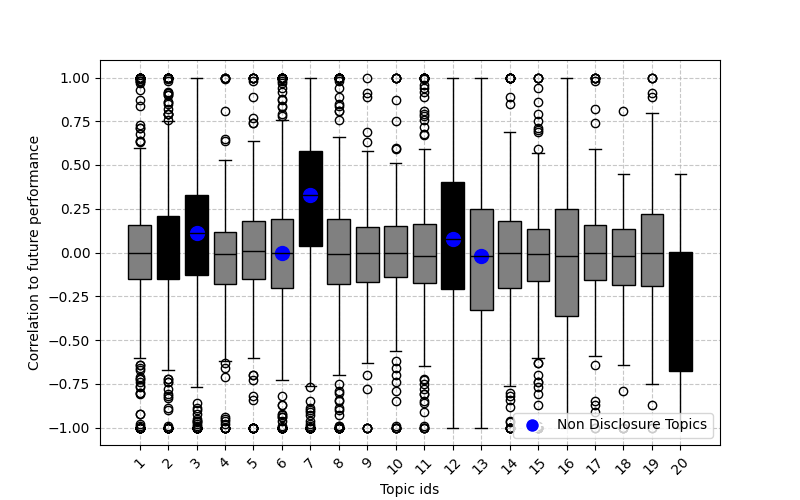}
            \caption{LDA 20}
        \end{subfigure}
        \hfill
        \begin{subfigure}{0.45\textwidth}
            \centering
            \includegraphics[width=\linewidth]{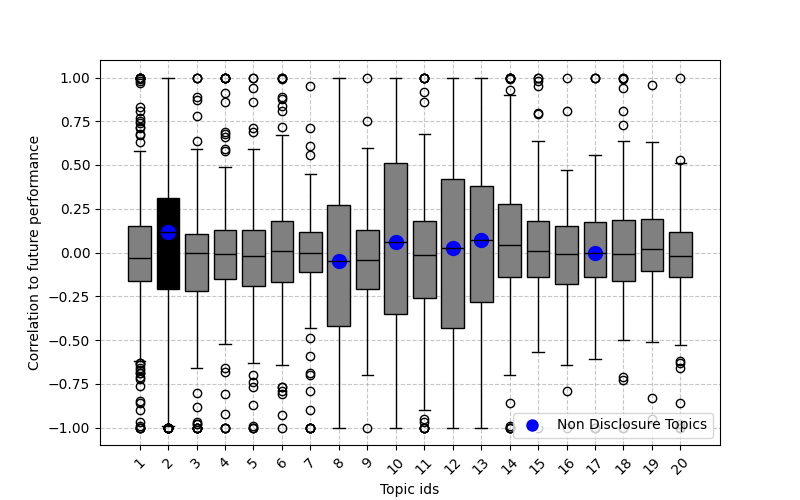}
            \caption{LDA 100}
        \end{subfigure}
                
        \begin{subfigure}{0.45\textwidth}
            \centering
            \includegraphics[width=\linewidth]{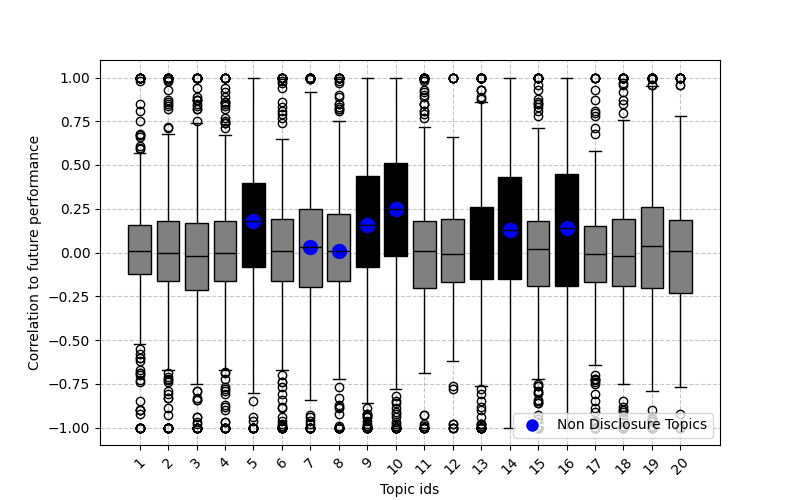}
            \caption{Top2Vec 20}
        \end{subfigure}
        \hfill
        \begin{subfigure}{0.45\textwidth}
            \centering
            \includegraphics[width=\linewidth]{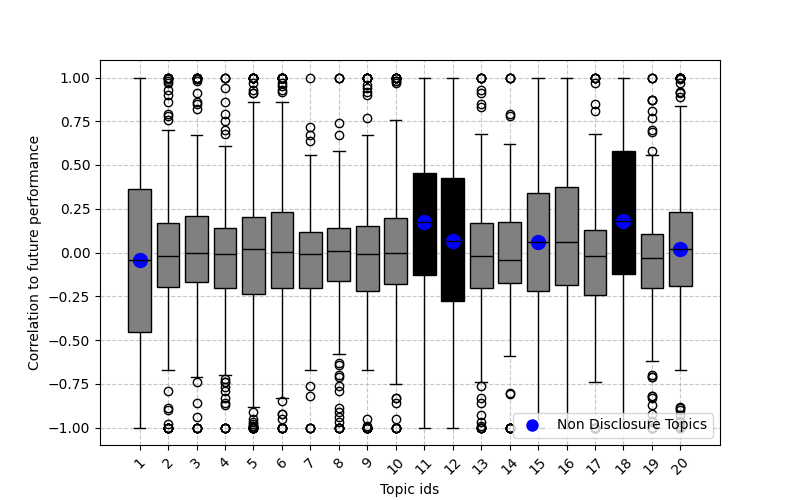}
            \caption{Top2Vec 100}
        \end{subfigure}
        
        \caption{Boxplots of the correlation between fund performance and topic sentiment scores generated using DistilBERT. Correlations are calculated at the individual fund level for each topic. Topics with bars colored in black demonstrate statistically significant mean correlation values (different from zero with p < 0.05), indicating strong evidence that sentiment expressed in these particular topics has predictive power for subsequent fund performance. Boxplots marked with blue dots in their center represent non-Disclosure topics (such as Market Update, Performance Commentary, and Strategy Overview), while unmarked boxplots correspond to Disclosure topics.}
        \label{fig:sentiment_score_distilbert}
    \end{figure}

\begin{figure}[ht]
        \centering
        \begin{subfigure}{0.45\textwidth}
            \centering
            \includegraphics[width=\linewidth]{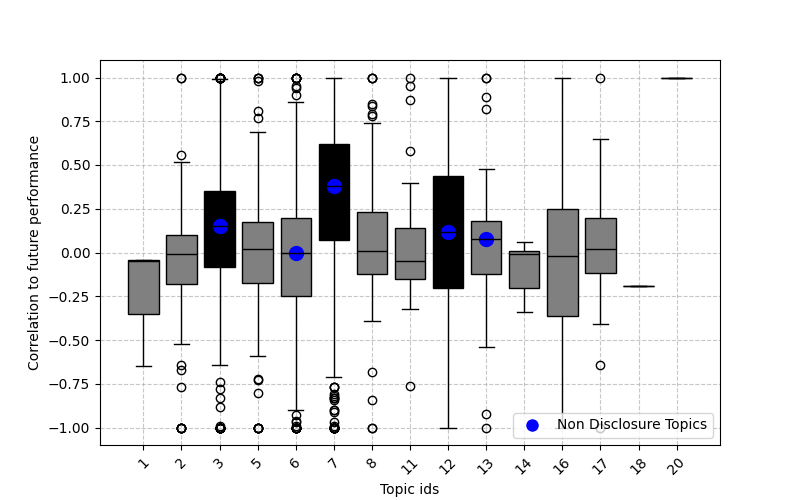}
            \caption{correlation analysis LDA 20}
        \end{subfigure}
        \hfill
        \begin{subfigure}{0.45\textwidth}
            \centering
            \includegraphics[width=\linewidth]{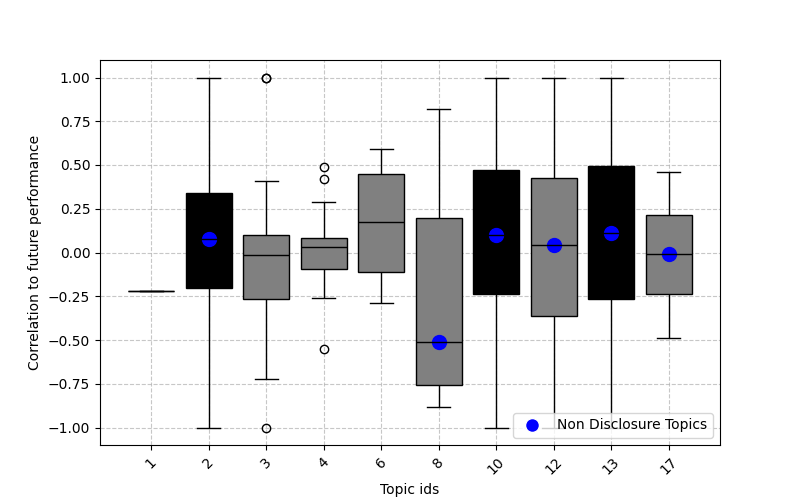}
            \caption{correlation analysis LDA 100}
        \end{subfigure}
                
        \begin{subfigure}{0.45\textwidth}
            \centering
             \includegraphics[width=\linewidth]{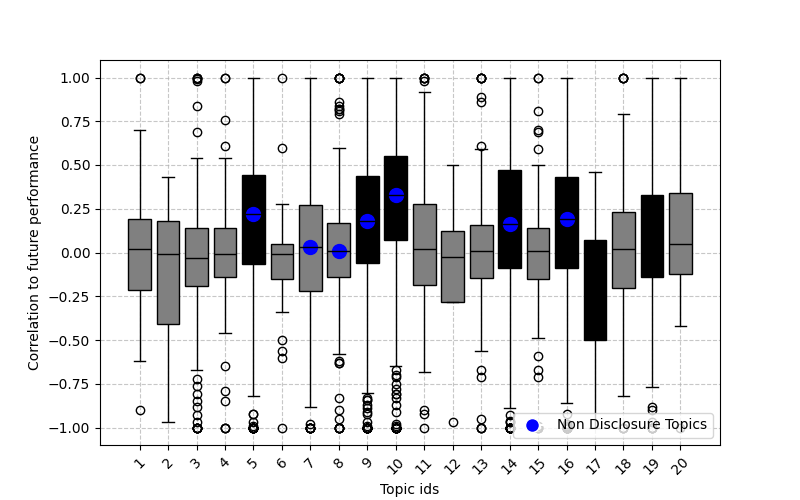}
            \caption{correlation analysis Top2Vec 20}
        \end{subfigure}
        \hfill
        \begin{subfigure}{0.45\textwidth}
            \centering
            \includegraphics[width=\linewidth]{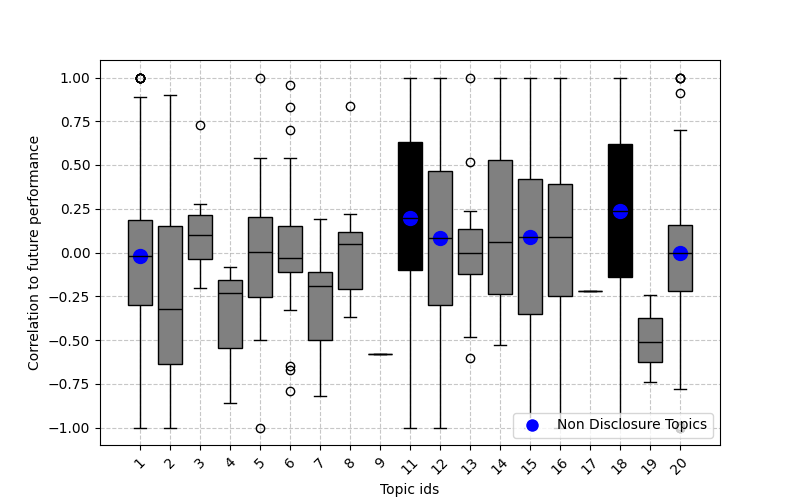}
            \caption{correlation analysis Top2Vec 100}
        \end{subfigure}
        
        \caption{Boxplots of the correlation between fund performance and topic sentiment scores generated using FinBERT. Correlations are calculated at the individual fund level for each topic. The boxplots display the maximum, minimum, and the 0.25, 0.50, and 0.75 quantiles of the correlations. Topics with bars colored in black demonstrate statistically significant mean correlation values (different from zero with p < 0.05), indicating strong evidence that sentiment expressed in these particular topics has predictive power for subsequent fund performance. Boxplots marked with blue dots in their center represent non-Disclosure topics (such as Market Update, Performance Commentary, and Strategy Overview), while unmarked boxplots correspond to Disclosure topics.}
        \label{fig:sentiment_score_finbert}
    \end{figure}

    The results of this analysis are illustrated in Figures \ref{fig:sentiment_score_distilbert} and \ref{fig:sentiment_score_finbert}, where we present correlation values for the first 20 topics generated by LDA 20, LDA 100, Top2Vec 20, and Top2Vec 100. These figures highlight topics with statistically significant correlations (p-value $<$ 0.05) in black, suggesting that they exhibit a meaningful relationship with fund performance. Table \ref{tab:sentiment_summary} provides a detailed statistical summary, including topics that are not labeled as "Disclosure" or those that exhibit significantly nonzero correlations. For each topic, we report the number of data points (i.e., sentiment-fund correlation scores per topic), the mean and standard deviation of correlation values, and the corresponding p-value, indicating whether the correlation is statistically significant. 
    
     Market Update and Performance Commentary exhibit the strongest correlations with fund performance among the annotated topics. These results suggest that sentiment in these topics contains valuable predictive information, likely because they reflect market sentiment and company performance evaluations. Conversely, Strategy Overview consistently shows weak and statistically insignificant correlations, indicating that high-level strategic discussions do not provide a reliable sentiment signal for predicting fund movements. Investment Team annotations yield inconsistent results, suggesting that while investment team expertise may hold value, their sentiment does not consistently align with future fund performance. 
    
    A key finding of our study is the performance gap between DistilBERT and FinBERT in generating sentiment scores. FinBERT shows stronger correlations with lower p-values where it performs well, but it often returns NaN (Not a Number) values, especially for text that are not financial news. This issue is prominent in sections like "Disclosure", "Investment Team", and "Strategy Overview", where many documents lack valid sentiment scores. While FinBERT performs comparably to DistilBERT on "Market Update" topics, it has a much higher rate of missing data in other categories. This suggests FinBERT’s training on financial news creates a bias: it excels with similar text but struggles with content using different vocabulary and structure. The results underscore a trade-off between domain specialization and generalizability in applying language models to varied financial documents.

    Based on DistilBERT, Top2Vec 20 consistently produces the strong correlations in topics not labeled as Disclosure, making it the most effective method for capturing sentiment-performance relationships.
    
    \begin{table}[]
        \centering
        \tiny
        \begin{tabular}{llcccccccc}
            \toprule
            & & \multicolumn{4}{c}{\textbf{DistilBert}} & \multicolumn{4}{c}{\textbf{Finbert}} \\
            \cmidrule(lr){3-6} \cmidrule(lr){7-10}
            \textbf{Topic id} & \textbf{Manual Annotation} & \textbf{count} & \textbf{mean} & \textbf{std} & \textbf{p\_value} & \textbf{count} & \textbf{mean} & \textbf{std} & \textbf{p\_value} \\
            \midrule
            \multicolumn{10}{l}{\textbf{LDA 20}} \\
            2  & Performance Commentary & 537 & 0.036 & 0.385  & \cellcolor{gray!30} 0.031  & 120  & -0.052 & 0.347 & 0.101 \\
            3  & Market Update          & 524 & 0.089 & 0.44 & \cellcolor{gray!30} 0.000        & 518  & 0.134  & 0.414 & \cellcolor{gray!30}0.000       \\
            6  & Strategy Overview      & 528 & -0.014  & 0.393 & 0.414  & 261  & -0.027 & 0.432 & 0.320  \\
            7  & Market Update          & 515 & 0.254 & 0.477 & \cellcolor{gray!30} 0.000        & 510  & 0.290  & 0.489 & \cellcolor{gray!30}0.000       \\
            8  & Performance Commentary & 289 & 0.001 & 0.448  & 0.957  & 51   & 0.062  & 0.458 & 0.337  \\
            12 & Performance Commentary & 323 & 0.071 & 0.500 & \cellcolor{gray!30} 0.011  & 305  & 0.100  & 0.507 & \cellcolor{gray!30}0.001 \\
            13 & Investment Team        & 217 & -0.026 & 0.57  & 0.500  & 27   & 0.050     & 0.502 & 0.609 \\
            20 & Disclosure             & 12  & -0.328 & 0.490 & \cellcolor{gray!30} 0.041  & 1    & 1        & nan     & nan     \\
            \midrule
            \multicolumn{10}{l}{\textbf{LDA 100}} \\
            2  & Market Update          & 369 & 0.063 & 0.486 & \cellcolor{gray!30}0.014  & 369  & 0.066   & 0.478 & \cellcolor{gray!30} 0.008 \\
            8  & Investment Team        & 145 & -0.045 & 0.627 & 0.384 & 14   & -0.252 & 0.611 & 0.146 \\
            10  & Market Update          & 261 & 0.046 & 0.613 & 0.225  & 259  & 0.080  & 0.593 & \cellcolor{gray!30}0.031 \\
            12 & Market Update          & 308 & 0.003  & 0.613 & 0.941  & 298  & 0.035   & 0.604 & 0.321 \\
            13 & Performance Commentary & 193 & 0.048 & 0.534 & 0.216  & 192  & 0.091  & 0.566 & \cellcolor{gray!30}0.0269 \\
            17 & Strategy Overview      & 91  & -0.001 & 0.356 & 0.983 & 16   & -0.016 & 0.260 & 0.806 \\
            \midrule
            \multicolumn{10}{l}{\textbf{Top2Vec 20}} \\
            5  & Market Update          & 522 & 0.145 & 0.431 & \cellcolor{gray!30} 0.00        & 515  & 0.196  & 0.431 & \cellcolor{gray!30}0.00       \\
                        7  & Investment Team      & 462 & 0.027 & 0.465 & 0.214  & 242  & 0.030   & 0.464 & 0.317 \\
            8  & Strategy Overview      & 547 & 0.014 & 0.383 & 0.371  & 232  & 0.003   & 0.367 & 0.898 \\
            9  & Market Update          & 535 & 0.147 & 0.4599  & \cellcolor{gray!30}0.00        & 528  & 0.160  & 0.459 & \cellcolor{gray!30}0.000       \\
            10  & Market Update          & 597 & 0.213 & 0.453 & \cellcolor{gray!30}0.000        & 558  & 0.274  & 0.444 & \cellcolor{gray!30}0.000       \\
            13 & Disclosure & 364 & 0.057 & 0.457 & \cellcolor{gray!30}0.018  & 91   & 0.027  & 0.416 & 0.536 \\
            14 & Performance Commentary & 430 & 0.123 & 0.502 & \cellcolor{gray!30}0.000        & 410  & 0.160  & 0.508 & \cellcolor{gray!30}0.000       \\
            16 & Market Update          & 406 & 0.108 & 0.524  & \cellcolor{gray!30}0.000  & 390  & 0.155  & 0.515 & \cellcolor{gray!30}0.000       \\
            17 & Disclosure             & 335 & -0.033 & 0.372 & 0.104  & 17   & -0.261 & 0.413 & \cellcolor{gray!30}0.019 \\
            19 & Disclosure             & 444 & 0.037 & 0.466 & 0.095  & 164  & 0.112  & 0.439 & \cellcolor{gray!30}0.001 \\
            \midrule
            \multicolumn{10}{l}{\textbf{Top2Vec 100}} \\
            1  & Investment Team        & 247 & -0.009 & 0.618 & 0.825  & 76   & -0.010 & 0.516 & 0.868 \\
            11 & Market Update          & 284 & 0.138 & 0.565 & \cellcolor{gray!30}0.000  & 282  & 0.188   & 0.569 & \cellcolor{gray!30}0.000       \\
            12 & Market Update          & 298 & 0.067 & 0.546 & \cellcolor{gray!30}0.034  & 290  & 0.048    & 0.566 & 0.150 \\
            15 & Business Update        & 253 & 0.060 & 0.499 & 0.056  & 169  & 0.028  & 0.556 & 0.513 \\
            18 & Strategy Overview      & 360 & 0.184 & 0.547 & \cellcolor{gray!30}0.000        & 353  & 0.174  & 0.584 & \cellcolor{gray!30}0.000       \\
            20 & Fund Overview          & 209 & 0.034 & 0.478 & 0.300  & 68   & -0.037 & 0.444 & 0.499 \\
            \bottomrule
        \end{tabular}
        \caption{Statistical summary of sentiment scores across different topic models. The table reports key metrics, including mean, standard deviation, and significance levels, to compare sentiment distributions among LDA 20, LDA 100, Top2Vec 20, and Top2Vec 100. Notably, Top2Vec 20 with DistilBERT demonstrates statistically significant correlations (highlighted in gray, p < 0.05) for all Market Update topics (topics 5, 9, 10, and 16) and Performance Commentary topics (topic 14), while only one Disclosure topic (topic 13) shows a slightly positive significant correlation. This consistent pattern suggests that Top2Vec 20 combined with DistilBERT is particularly effective at identifying topics with sentiment values that correlate with future fund performance.}
        \label{tab:sentiment_summary}
    \end{table}

        \section{Conclusion}
    \label{outlook}
    This study investigates the effectiveness of advanced topic modeling techniques in analyzing hedge fund documents and their potential applications in financial decision-making. We compare Latent Dirichlet Allocation (LDA), Top2Vec, and BERTopic, evaluating their ability to identify meaningful topics and assessing their robustness across different configurations. Additionally, we examine the relationship between sentiment and future fund performance by leveraging both a general-purpose sentiment model (DistilBERT) and a domain-specific financial model (FinBERT).

    Our findings indicate that LDA provides the most interpretable and reliable topic classifications, particularly for human users. Among the tested configurations, LDA 20 generates the most coherent topics, while Top2Vec shows superior classification performance. LDA 100 improves granularity but introduces higher variability. Despite utilizing word embeddings, BERTopic fail to outperform LDA in clustering hedge fund document text. We also assess the robustness of these topic models by analyzing their stability when varying the number of topics. The confusion matrix comparisons demonstrate that LDA exhibits greater consistency in topic assignment than Top2Vec, suggesting that LDA provides a more stable and reliable topic structure. In contrast, Top2Vec's topics shift significantly as the number of topics changes, reducing its reliability. Although the FinBERT is fine-tuned specifically with financial data, it turns out to be too specific in financial news, so it failed to generate sentiment scores for topics that are less financial news related. The combination of Top2Vec 20 and DistilBERT produces the strongest correlation, demonstrating its potential for automating hedge fund document analysis. These findings suggest that investors and analysts can enhance their decision-making processes by leveraging NLP-driven insights, thereby improving efficiency in processing unstructured hedge fund data.

    Future research could explore the implementation of ensemble models to address the limitations of pre-trained embeddings, which are not specifically tailored for hedge fund text. One potential approach is to apply LDA prior to Top2Vec to filter out disclosure text that is less relevant for investment analysis. Also, future work could extend this study by integrating supervised learning techniques to optimize and automate the topic categorization or by incorporating pre-trained models that are fine-tuned with our dataset.

\appendix
\section{Dataset and Data Cleaning}
    \label{appendix_data_preprocessing}
    The dataset used in this study comprises various types of fund-related documents, including fact sheets, presentations, quarterly reports, monthly reports, and investor letters. The classification of document types follows common industry practices and was manually determined. All documents were provided in PDF format. To extract the text at the block (paragraph) level, we employed docling \citep{docling} as an Optical Character Recognition (OCR) tool. The OCR process identified approximately 1.02 million text blocks containing at least five words. For the correlation analysis, monthly performance data was extracted manually from the hedge fund documents.

    The extracted paragraph text underwent several preprocessing steps. First, all non-English content was removed. Next, paragraphs containing fewer than 50 words were discarded to eliminate incomplete or uninformative text. To prevent text truncation during embedding, paragraphs exceeding 400 words were segmented into overlapping chunks, with each chunk having a maximum length of 400 words and an overlap of 50 words between consecutive chunks. Some of the example text can be found in table \ref{tab:example_text}

    For LDA-based topic modeling, additional preprocessing was applied to the text corpus. This included:
    \begin{enumerate}
        \item Converting text to lowercase to ensure uniformity,
        \item Removing common stop words to reduce noise,
        \item Applying part-of-speech (POS)-based lemmatization to standardize word forms,
        \item Eliminating very short words that do not contribute to meaningful topic extraction.
    \end{enumerate}

    \begin{landscape}
    \begin{table}
    \tiny
        \centering
\begin{tabular}{|p{18cm}|}
\hline
\textbf{Paragraph} \\
\hline
The information and opinions contained herein have been compiled or arrived at in good faith based upon information obtained from
sources believed to be reliable. However, such information has not been independently verified and no guarantee, representation or
warranty, express or implied, is made as to its accuracy, completeness or correctness. All such information and opinions are
subject to change without notice. Descriptions of entities and securities mentioned herein are not intended to be complete. This
document is for information purposes only. This document is not, and should not be construed as, an offer, or solicitation of an
offer, to buy or sell any securities or other financial instruments. The investment program described herein is offered solely on
the basis of the information and representations expressly set forth in the relevant offering circulars, and no other information
or representations may be relied upon in connection with the offering of certificate. Performance information is not a measure of
return to the investor, is not based on audited financial statements, and is dated; return may have decreased since the issuance
of this report. Past performance is not necessarily an indication for future results. This document may not be reproduced,
distributed, published or delivered to any other party for any purposes. \\
\hline
Information for Investors in Switzerland: The offering of the Shares in Switzerland will be exclusively made to, and directed at,
qualified investors (the 'Qualified Investors'), as defined in Article 10 para. 3 and 3ter of the Swiss Collective Investment
Schemes Act of 23 June 2006, as amended from time to time ('CISA'), to the exclusion of any high-networth individuals and their
private investment structures with an opting-out pursuant to the Article 5 para. 1 of the Swiss Financial Services Act ('FinSA').
The Fund has not been and will not be approved by the Swiss Financial Market Supervisory Authority ('FINMA') for offering to non-
qualified investors. Therefore, the Explanatory Memorandum, any other offering materials and/or marketing materials relating to
the Shares may be made available in Switzerland solely to Qualified Investors. \\
\hline
\{...\} operates its investment strategy through a master-feeder structure by
investing substantially all of its assets in \{...\}. Information provided herein
relating to the investments and assets under management of the Fund includes investments made through the Master Fund. Prior to
\{...\}, investors subscribed for shares directly in the Master Fund. \\
\hline
Many of the bonds issued during this period remain materially out of the money and are unlikely to be converted into equity. As a
result, this debt will need to be repaid at maturity. If this debt is trading significantly below \{...\}, a liability will
be owed. We believe it is sensible for companies to repurchase these bonds at a discount to \{...\} and potentially issue new
convertible securities. \\
\hline
Monthly Review In May, global financial markets were buffeted by two divergent US data surprises. Early in the month, non-farm
payrolls shocked massively to the downside, while a few days later core and headline consumer price inflation surpassed already-
elevated expectations. A hawkish reconciliation of these data points would argue that enhanced unemployment benefits and pickier
jobseekers constrained labour supply in a manner which will ultimately fuel wage-push inflation. A dovish interpretation would
reply that these data points are consistent with the Federal Reserve's professed strategy, in which it can afford to be patient in
the face of transient inflation, while the economy heals from the pandemic. \\
\hline
The reinvestment of dividends, if any. Differences in account size, timing of transactions and market conditions prevailing at the
time of investment may lead to different results. Differences in the methodology used to calculate performance may also lead to
different performance results than those shown. Broad-based securities indices are unmanaged and are not subject to fees and
expenses typically associated with managed accounts or investment funds. Investments cannot be made directly in a broad-based
securities index. \\
\hline
Month in Review The first month of 2024 brought strong real economic data in the U.S., calling into question the extraordinary
rates rally at yearend. The Q4 2023 GDP came out much higher-than-expected at 3.3\% quarter-on-quarter annualised, and the Atlanta
Fed's GDP Now Forecast for the present quarter suggested little deceleration from that pace. The unemployment rate dropped 0.1\% to
3.7\%, retail sales and University of Michigan sentiment came out very strong, and CPI surprised slightly to the high side. As
such, it was not surprising that Federal Reserve Chair \{...\} changed his tone in the meeting on the final day of the month,
suggesting that the \{...\} wanted to see further evidence of inflation's downward path before commencing rate cuts. Unexpectedly, he
explicitly stated that a cut in the March meeting was unlikely. One more subtle hint of his shift in bias was his emphasis on the
12-month annualised measure of inflation, which remains well above target, rather than the 6-month annualised number mentioned in
the December press conference, which is much lower. \{...\}'s change in tone contrasted with
\{...\}'s pivot the previous week. In the December meeting she argued that the ECB wanted to see wage data available only by the
middle of the year before deciding to cut rates. However, in the January meeting she argued that the ECB would have earlier
indications of wage pressure, opening the door to earlier rate cuts. Ultimately, the strong U.S. data and Fed shift left yields
higher on the month, but the scale of the change was moderated by a risk-off scare on the same day as the Fed meeting. \{...\}, a relatively small lender, reported losses attributed to commercial real estate. This news awakened fears of
systemic risk, just as \{...\}'s problems did last spring. The December 2024 3-month SOFR contract, a rough measure of
where the U.S. policy rate is expected to be at the end of the year, ended the month only 9.5 basis points higher than where it
was at the end of December. \\
\hline
Risk Management Framework (2 of 2) 19 o Investment risk limits o Portfolio size as defined by \{...\} o Net market exposure (dollar
and beta) o Gross and net sector/industry exposures o Gross and net geographical limits o Position concentration limits (single
position, top 5 positions) o Minimum number of positions o Gross and net exposures within market capitalization ranges o Gross and
net exposures within liquidity ranges o Gross and net exposures within options positions o Maximum ETF and futures exposure o
Drawdown limits o Incremental portfolio allocation reductions based on portfolio allocation thresholds (peak-to-trough)
Investment/Drawdown Risk Limits Portfolio Exposure Management in Real Time · \{...\}'s risk management team leverages off-the-
shelf and proprietary risk factors to manage portfolio exposures in real-time across the portfolio. · Datapoints are continually
collected across a wide breadth of sources and internally analyzed to identify tangible factors to assist with distinguishing
sources of performance drivers vs. risk exposures. · Data \& market intelligence sources: · Vendor sources · Exchange-traded
vehicles · Pure long/short factors · Regulatory filings · Sell-side data · Internal data · Market / industry structure
(macro/micro) · Behavioral observations · Flow data · Environmental indicators Ref [ \{...\} ] \\
\hline

\hline
\end{tabular}
        \caption{Some example text from dataset, sensitive data is replaced with \{...\}}
        \label{tab:example_text}
    \end{table}
    \end{landscape}

    \section{Impact of n-gram}
    \label{appendix_ngram}
    To evaluate the impact of n-gram modeling on topic generation, we experimented with two variations of Top2Vec and BERTopic: one utilizing only uni-gram (Top2Vec and BERTopic) and another incorporating n-gram (Top2Vec ngram and BERTopic ngram), which include phrases of one, two, and three words. However, the distribution of documents across topics remained almost unchanged between the uni-gram and n-gram implementations. Given the minimal impact of n-gram modeling on topic structure, we opted to use the uni-gram models for model validation to simplify analysis.

    \section{Tables of Topics}
    \label{appendix_topic_models_tables}

    \begin{landscape}
    \begin{table}
    \scriptsize
        \centering
        \begin{tabular}{lcrcc}
        \hline
            No. & Top-10 keywords & \#paragraphs & LLM \& Manual Annotation\\ 
                \hline

            Topic 1 & \makecell{information, document, contain, investment, make, confidential, offer,\\ advice, memorandum, provide} & 54,879 & Disclosure \\
            Topic 2 & \makecell{fee, performance, return, net, expense, class, fund, management, incentive, allocation} & 41,117 &  Disclosure \\
            Topic 3 & \makecell{inflation, rate, market, year, growth, policy, economy, fed, remain, continue} & 38,325 & Market Update \\
            Topic 4 & \makecell{law, security, offer, person, regulation, jurisdiction, public, offering, sell, act} & 25,870 &  Disclosure \\
            Topic 5 & \makecell{risk, investment, fund, involve, loss, substantial, investor, invest, speculative, degree} & 25,055 & Disclosure \\
            Topic 6 & \makecell{strategy, portfolio, fund, opportunity, manager, arbitrage, long, focus, market, alpha} & 23,784 & Strategy Overview \\
            Topic 7 & \makecell{month, position, short, long, gain, positive, equity, sector, oil, contribute} & 22,179 &  Performance Commentary \\
            Topic 8 & \makecell{exposure, calculate, value, portfolio, position, gross, ratio, return, aum, percentage} & 18,001 & Disclosure \\
            Topic 9 & \makecell{singapore, sfa, australian, australia, authority, section, register, corporation, offer, uae} & 17,621 &  Disclosure \\
            Topic 10 & \makecell{switzerland, swiss, representative, office, agent, fund, registered, kuwait, ireland, pay} & 17,517 & Disclosure \\
            Topic 11 & index, benchmark, msci, comparison, hfri, bloomberg, weight, fund, market, performance & 15,597 & Disclosure\\
            Topic 12 & \makecell{company, deal, business, quarter, announce, revenue, new, believe, large, digital} & 15,392 & Performance Commentary \\
            Topic 13 & \makecell{join, firm, university, capital, management, head, howard, chief, officer, brevan} & 13,193 & Investment Team \\
            Topic 14 & \makecell{statement, forward, look, assumption, result, actual, estimate, future, materially, projection} & 13,142 & Disclosure\\
            Topic 15 & \makecell{act, person, financial, united, kingdom, conduct, fca, order, promotion, service} & 11,499 & Disclosure\\
            Topic 16 & \makecell{credit, bond, loan, debt, yield, corporate, spread, clo, market, value} & 8,184 &  other \\
            Topic 17 & \makecell{trading, program, hypothetical, sfef, result, account, forma, sfeof, commodity, pro} & 7,378 &Disclosure\\
            Topic 18 & \makecell{state, member, aifmd, directive, eea, lawfully, implement, initiative, place, alternative} & 7,365 & Disclosure \\
            Topic 19 & \makecell{kong, hong, ordinance, amundi, professional, asia, content, gips, exercise, rule} & 4,739 &  Disclosure \\
            Topic 20 & \makecell{font, cidfont, glyph, sspf, timesnewroman, lcklih, sspof, arial, lclejb, raleway} & 638 &  Disclosure \\
                    \hline

        \end{tabular}
        \caption{Sample topics generated by LDA with 20 classes and 50 iterations. The table presents the top words for each topic, along with the number of assigned paragraph text and representative annotations.}
        \label{tab:topics_lda_20}
    \end{table}
    \end{landscape}

    \begin{landscape}
    \begin{table}
    \tiny
        \centering
            \begin{tabular}{lccccccclc}
            \hline
            No. & Disclosure & Fund Terms & Investment Team & Market Update & Performance Commentary & Strategy Overview & other & Predicted Class & Accuracy \\ \hline
            Topic 1 & 99.92 & 0.07 & 0.00 & 0.00 & 0.00 & 0.00 & 0.01 & Disclosure & 99.92 \\ 
            Topic 2 & 71.37 & 13.14 & 0.00 & 0.00 & 13.25 & 0.83 & 1.41 & Disclosure & 71.37 \\ 
            Topic 3 & 0.65 & 0.00 & 0.07 & 79.65 & 10.38 & 6.69 & 2.56 & Market Update & 79.65 \\ 
            Topic 4 & 98.91 & 1.08 & 0.00 & 0.00 & 0.00 & 0.00 & 0.01 & Disclosure & 98.91 \\ 
            Topic 5 & 96.71 & 1.57 & 0.01 & 0.10 & 0.40 & 0.83 & 0.38 & Disclosure & 96.71 \\ 
            Topic 6 & 3.79 & 1.55 & 3.71 & 0.11 & 2.01 & 71.58 & 17.25 & Strategy Overview & 71.58 \\ 
            Topic 7 & 0.46 & 0.00 & 0.00 & 16.04 & 79.16 & 3.65 & 0.69 & Performance Commentary & 79.16 \\ 
            Topic 8 & 77.75 & 1.47 & 0.18 & 0.10 & 8.02 & 4.15 & 8.34 & Disclosure & 77.75 \\ 
            Topic 9 & 99.78 & 0.03 & 0.17 & 0.00 & 0.00 & 0.01 & 0.01 & Disclosure & 99.78 \\ 
            Topic 10 & 99.56 & 0.25 & 0.02 & 0.00 & 0.00 & 0.00 & 0.16 & Disclosure & 99.56 \\ 
            Topic 11 & 81.53 & 0.81 & 0.00 & 2.46 & 2.19 & 0.26 & 12.75 & Disclosure & 81.53 \\ 
            Topic 12 & 2.78 & 0.20 & 1.51 & 13.34 & 40.69 & 21.15 & 20.32 & Performance Commentary & 40.69 \\ 
            Topic 13 & 9.49 & 0.03 & 87.95 & 0.00 & 0.79 & 0.03 & 1.71 & Investment Team & 87.95 \\ 
            Topic 14 & 99.95 & 0.00 & 0.00 & 0.05 & 0.00 & 0.00 & 0.00 & Disclosure & 99.95 \\ 
            Topic 15 & 100.00 & 0.00 & 0.00 & 0.00 & 0.00 & 0.00 & 0.00 & Disclosure & 100.00 \\ 
            Topic 16 & 27.14 & 2.82 & 1.50 & 13.56 & 7.96 & 18.39 & 28.64 & other & 28.64 \\ 
            Topic 17 & 78.63 & 0.00 & 0.58 & 0.00 & 0.81 & 18.80 & 1.18 & Disclosure & 78.63 \\ 
            Topic 18 & 100.00 & 0.00 & 0.00 & 0.00 & 0.00 & 0.00 & 0.00 & Disclosure & 100.00 \\ 
            Topic 19 & 99.51 & 0.00 & 0.00 & 0.00 & 0.00 & 0.11 & 0.38 & Disclosure & 99.51 \\ 
            Topic 20 & 73.46 & 1.85 & 0.00 & 1.23 & 7.41 & 6.79 & 9.26 & Disclosure & 73.46 \\ 
            \hline
            \end{tabular}
            \caption{Classification accuracy and content distribution for topics identified by the LDA 20 model applied to hedge fund text. This table presents the percentage breakdown of text samples within each topic across six predefined document categories, plus an "other" category for content that doesn't fit these classifications. Each row represents one of the 20 topics extracted by LDA 20, while columns represent the document categories: "Disclosure" (legal disclaimers and regulatory statements), "Fund Terms" (fee structures and investment conditions), "Investment Team" (personnel information), "Market Update" (economic and market commentary), "Performance Commentary" (fund results and attribution), and "Strategy Overview" (investment approach and philosophy). The "Predicted Class" column indicates the dominant category for each topic based on the highest percentage value in the row, while the "Accuracy" column shows the corresponding percentage of correctly classified text samples within that topic. For example, Topic 12 is classified as "Investment Team" with 87.95\% accuracy, meaning that nearly 88\% of text samples in this topic contain information about fund personnel. The results demonstrate that LDA effectively clusters semantically similar content, with many topics showing high classification accuracy (>90\% for Topics 0, 3, 4, 8, 9, 13, 14, 17, and 18). However, some topics exhibit mixed content (e.g., Topic 15) or moderate accuracy (40-80\%), indicating areas where classification is more challenging. This analysis was conducted using a combination of Large Language Model (ChatGPT) classification and manual review to ensure accuracy.}
        \label{tab:metrics_lda_20}
    \end{table}
    \end{landscape}
    
    \begin{landscape}
    \begin{table}
    \scriptsize
        \centering
        \begin{tabular}{lcrc}
                \hline

            No. & Top-10 keywords & \#paragraphs & LLM \& Manual Annotation\\ 
                    \hline

            Topic 1 & \makecell{offering, memorandum, offer, solicitation, document, make, fund, buy, sell, contain} & 12,464 & Disclosure\\
            Topic 2 & \makecell{inflation, rate, fed, cut, hike, central, bank, rise, policy, market} & 10,218 & Market Update\\
            Topic 3 & \makecell{risk, investment, fund, speculative, substantial, involve, degree, investor, loss, leverage} & 8,610 & Disclosure\\
            Topic 4 & \makecell{fee, incentive, return, expense, performance, net, management, allocation, investor, timing} & 8,136 & Disclosure\\
            Topic 5 & \makecell{kingdom, financial, promotion, order, person, fca, united, service, conduct, act} & 7,488 &  Disclosure\\
            Topic 6 & \makecell{index, benchmark, comparison, unmanaged, fund, comprise, performance, security, invest, comparative} & 7,188 & Disclosure\\
            Topic 7 & \makecell{switzerland, representative, swiss, office, registered, geneva, agent, pay, zurich, banque} & 7,052 &  Disclosure\\
            Topic 8 & \makecell{join, head, university, chief, officer, committee, school, director, executive, prior} & 6,645 &  Investment Team\\
            Topic 9 & \makecell{singapore, sfa, section, canada, invitation, offer, prospectus, pursuant, security, canadian} & 6,538 &  Disclosure\\
            Topic 10 & \makecell{china, covid, election, economy, fiscal, policy, chinese, ukraine, government, stimulus} & 6,437 &  Market Update\\
            Topic 11 & \makecell{statement, forward, look, terminology, uncertainty, actual, event, materially, variation, contemplate} & 6,166 &  Disclosure\\
            Topic 12 & \makecell{market, environment, think, year, remain, like, believe, opportunity, continue, just} & 6,048 &  Market Update\\
            Topic 13 & \makecell{quarter, company, continue, position, attractive, strong, restructuring, cash, believe, remain} & 5,811 &  Performance Commentary\\
            Topic 14 & \makecell{advice, legal, tax, accounting, investment, recommendation, consult, financial, constitute, construe} & 5,721 & Disclosure\\
            Topic 15 & \makecell{confidential, reproduce, write, consent, recipient, strictly, information, disclose, copy, person} & 5,657 &  Disclosure\\
            Topic 16 & \makecell{state, member, lawfully, implement, aifmd, initiative, place, distribute, relevant, eea} & 5,277 &  Disclosure\\
            Topic 17 & \makecell{trend, systematic, strategy, global, program, diversified, market, diversify, traditional, contrarian} & 5,177 & Strategy Overview\\
            Topic 18 & \makecell{act, state, united, security, amend, purchaser, register, define, company, person} & 4,828 & Disclosure\\
            Topic 19 & \makecell{korea, japan, law, japanese, israel, sophisticated, joint, addendum, resident, nis} & 4,409 &  Disclosure\\
            Topic 20 & \makecell{jurisdiction, document, regulation, distribution, inform, law, person, responsibility, country, restrict} & 4,387 &  Disclosure\\

                    \hline

        \end{tabular}
        \caption{Sample topics generated by LDA with 100 classes and 50 iterations. The table presents the top words for each topic, along with the number of assigned paragraph text and representative annotations.}
        \label{tab:topics_lda_100}
    \end{table}
    \end{landscape}

    \begin{landscape}
    \begin{table}
    \tiny
        \centering
            \begin{tabular}{lccccccclc}
            \hline
            No. & Disclosure & Fund Terms & Investment Team & Market Update & Performance Commentary & Strategy Overview & other & Predicted Class & Accuracy (\%) \\ \hline
            Topic 1 & 99.93 & 0.07 & 0.00 & 0.00 & 0.00 & 0.00 & 0.00 & Disclosure & 99.93 \\ 
            Topic 2 & 0.00 & 0.00 & 0.00 & 93.31 & 5.43 & 1.11 & 0.14 & Market Update & 93.31 \\ 
            Topic 3 & 99.62 & 0.05 & 0.00 & 0.00 & 0.00 & 0.18 & 0.15 & Disclosure & 99.62 \\ 
            Topic 4 & 79.78 & 12.84 & 0.00 & 0.00 & 7.25 & 0.06 & 0.06 & Disclosure & 79.78 \\ 
            Topic 5 & 100.00 & 0.00 & 0.00 & 0.00 & 0.00 & 0.00 & 0.00 & Disclosure & 100.00 \\ 
            Topic 6 & 99.79 & 0.00 & 0.00 & 0.03 & 0.12 & 0.03 & 0.03 & Disclosure & 99.79 \\ 
            Topic 7 & 99.41 & 0.59 & 0.00 & 0.00 & 0.00 & 0.00 & 0.00 & Disclosure & 99.41 \\ 
            Topic 8 & 0.11 & 0.00 & 99.71 & 0.00 & 0.06 & 0.00 & 0.11 & Investment Team & 99.71 \\ 
            Topic 9 & 100.00 & 0.00 & 0.00 & 0.00 & 0.00 & 0.00 & 0.00 & Disclosure & 100.00 \\ 
            Topic 10 & 3.75 & 0.00 & 0.00 & 85.18 & 3.36 & 1.19 & 6.52 & Market Update & 85.18 \\ 
            Topic 11 & 100.00 & 0.00 & 0.00 & 0.00 & 0.00 & 0.00 & 0.00 & Disclosure & 100.00 \\ 
            Topic 12 & 0.00 & 0.00 & 0.20 & 47.52 & 18.61 & 30.10 & 3.56 & Market Update & 47.52 \\ 
            Topic 13 & 0.17 & 0.50 & 0.34 & 5.36 & 68.51 & 20.77 & 4.36 & Performance Commentary & 68.51 \\ 
            Topic 14 & 100.00 & 0.00 & 0.00 & 0.00 & 0.00 & 0.00 & 0.00 & Disclosure & 100.00 \\ 
            Topic 15 & 100.00 & 0.00 & 0.00 & 0.00 & 0.00 & 0.00 & 0.00 & Disclosure & 100.00 \\ 
            Topic 16 & 100.00 & 0.00 & 0.00 & 0.00 & 0.00 & 0.00 & 0.00 & Disclosure & 100.00 \\ 
            Topic 17 & 0.18 & 0.00 & 1.49 & 0.00 & 0.18 & 88.77 & 9.38 & Strategy Overview & 88.77 \\ 
            Topic 18 & 99.91 & 0.09 & 0.00 & 0.00 & 0.00 & 0.00 & 0.00 & Disclosure & 99.91 \\ 
            Topic 19 & 100.00 & 0.00 & 0.00 & 0.00 & 0.00 & 0.00 & 0.00 & Disclosure & 100.00 \\ 
            Topic 20 & 100.00 & 0.00 & 0.00 & 0.00 & 0.00 & 0.00 & 0.00 & Disclosure & 100.00 \\ 
            \hline
            \end{tabular}
            \caption{Classification accuracy and content distribution for topics identified by the LDA 100 model applied to hedge fund text. This table presents the percentage breakdown of text samples within each topic across six predefined document categories, plus an "other" category for content that doesn't fit these classifications. Each row represents one of the 20 topics extracted by LDA 20, while columns represent the document categories: "Disclosure" (legal disclaimers and regulatory statements), "Fund Terms" (fee structures and investment conditions), "Investment Team" (personnel information), "Market Update" (economic and market commentary), "Performance Commentary" (fund results and attribution), and "Strategy Overview" (investment approach and philosophy). The "Predicted Class" column indicates the dominant category for each topic based on the highest percentage value in the row, while the "Accuracy" column shows the corresponding percentage of correctly classified text samples within that topic.}
            \caption{metrics of lda 100}
        \label{tab:metrics_lda_100}
    \end{table}
    \end{landscape}

        \begin{landscape}
    \begin{table}
    \scriptsize
        \centering
        \begin{tabular}{lcrc}
                \hline

            No. & Top-10 keywords & \#paragraphs & LLM \& Manual Annotation\\
                    \hline

            Topic 1 & \makecell{investor, securities, investorrights, investors, investees, fund, investments, invests, invest, theinvestment} & 41,364 &  Disclosure\\
            
            Topic 2 & \makecell{securities, investor, investorrights, investment, invest, investments, investible, investability,\\ invests, theinvestment} & 35,419 & Disclosure\\
            
            Topic 3 & \makecell{investments, investment, invest, investability, invests, investing, securities, investors, investable, investible} & 32,689 &  Disclosure\\
            
            Topic 4 & \makecell{investment, underinvestment, portfolios, financials, earnings, theinvestment, investments, portfolio,\\ invests, securities} & 31,940 &  Disclosure\\
            
            Topic 5 & \makecell{inflation, recessions, macroeconomic, recession, inflationary, recessionary, markets, treasury, forex, monetary} & 31,765 &  Market Update\\
            
            Topic 6 & \makecell{securities, investor, ltda, investorrights, investcorp, financials, theinvestment, investors, investm, financial} & 30,377 &  Disclosure\\
            
            Topic 7 & \makecell{investors, securities, investor, investments, theinvestment, investmen, portfolios, assetmanagement, investcorp, invests} & 27,936 & Investment Team \\
            
            Topic 8 & \makecell{investments, investment, invest, invests, portfolios, portfolio, investors, investing, fund, investor} & 26,171 & Strategy Overview\\
            
            Topic 9 & \makecell{recessions, markets, recession, investments, investing, recessionary, investors, stocks, investability, investment} & 24,402 & Market Update\\
            
            Topic 10 & \makecell{investments, investors, theinvestment, underinvestment, investing, investor, invests, investment,\\ invest, reinvestments} & 22,991 &  Performance Commentary\\
            
            Topic 11 & \makecell{indexes, index, indices, indexation, indexed, portfolios, invests, investments, investability, securities} & 22,910 & Disclosure\\

            Topic 12 & \makecell{securities, investor, investorrights, investors, noninvestment, theinvestment, investcorp,\\ invests, invest, investments} & 22,765 &  Disclosure \\

            Topic 13 & \makecell{securities, portfolios, assetmanagement, portfolio, investments, theinvestment, financials,\\ valuations, investment, underinvestment} & 22,692 & Disclosure\\

            Topic 14 & \makecell{reinvestments, theinvestment, reinvestment, investcorp, underinvestment, markets, market,\\ investors, liquidations, firms} & 22,062 & Performance Commentary\\

            Topic 15 & \makecell{investability, investments, investment, invests, invest, investible, investmentgrade, invested,\\ investing, investable} & 21,548 & Disclosure\\

            Topic 16 & \makecell{financings, theinvestment, finance, financing, reinvestments, financials, underinvestment, subinvestment,\\ leverages, reinvestment} & 21,318 & Market Update\\

            Topic 17 & \makecell{disclosures, disclosure, disclose, confidentiality, informing, confidentially, documents, authorisations,\\ securities, disclaimers} & 21,184 &  Disclosure\\

            Topic 18 & \makecell{investability, investments, futures, investment, prospects, profitability, investible, portfolios,\\ forecasting, investing} & 19,930 &  Disclosure\\

            Topic 19 & \makecell{investmentgrade, financials, investment, investments, etf, underinvestment, securities, invests,\\ nvestment, financial} & 19,762 &  Disclosure\\

            Topic 20 & \makecell{securities, investor, investorrights, investors, noninvestment, theinvestment, nvestment, invest, invests, investm} & 16,927 & Disclosure\\

                    \hline

        \end{tabular}
        \caption{Sample topics generated by Top2Vec after reducing the number of topics to 20. The table presents the top words for each topic, the number of assigned paragraph text, and representative annotations.}
        \label{tab:topics_top2vec}
    \end{table}
    \end{landscape}

        \begin{landscape}
    \begin{table}
    \tiny
        \centering
            \begin{tabular}{lccccccclc}
            \hline
            No. & Disclosure & Fund Terms & Investment Team & Market Update & Performance Commentary & Strategy Overview & other & Predicted Class & Accuracy (\%) \\ \hline
            Topic 1 & 99.25 & 0.46 & 0.00 & 0.02 & 0.03 & 0.04 & 0.21 & Disclosure & 99.25 \\ 
            Topic 2 & 99.40 & 0.13 & 0.00 & 0.01 & 0.00 & 0.25 & 0.21 & Disclosure & 99.40 \\ 
            Topic 3 & 91.25 & 0.96 & 0.01 & 0.22 & 0.02 & 3.67 & 3.88 & Disclosure & 91.25 \\ 
            Topic 4 & 76.11 & 7.30 & 0.00 & 0.01 & 9.12 & 3.50 & 3.95 & Disclosure & 76.11 \\ 
            Topic 5 & 4.57 & 0.00 & 0.00 & 66.25 & 25.60 & 2.73 & 0.85 & Market Update & 66.25 \\ 
            Topic 6 & 98.69 & 0.48 & 0.11 & 0.03 & 0.01 & 0.08 & 0.59 & Disclosure & 98.69 \\ 
            Topic 7 & 13.04 & 0.18 & 51.61 & 0.18 & 0.79 & 24.12 & 10.07 & Investment Team & 51.61 \\ 
            Topic 8 & 17.39 & 5.31 & 2.34 & 0.17 & 2.03 & 64.49 & 8.26 & Strategy Overview & 64.49 \\ 
            Topic 9 & 6.17 & 0.05 & 0.14 & 50.14 & 18.02 & 21.70 & 3.78 & Market Update & 50.14 \\ 
            Topic 10 & 3.95 & 0.61 & 0.09 & 6.01 & 68.38 & 18.55 & 2.41 & Performance Commentary & 68.38 \\ 
            Topic 11 & 76.71 & 0.69 & 0.00 & 2.60 & 1.98 & 1.19 & 16.83 & Disclosure & 76.71 \\ 
            Topic 12 & 98.62 & 1.14 & 0.00 & 0.03 & 0.03 & 0.00 & 0.19 & Disclosure & 98.62 \\ 
            Topic 13 & 73.88 & 4.09 & 0.31 & 0.18 & 4.53 & 4.85 & 12.15 & Disclosure & 73.88 \\ 
            Topic 14 & 2.36 & 0.28 & 0.51 & 19.19 & 46.04 & 19.10 & 12.52 & Performance Commentary & 46.04 \\ 
            Topic 15 & 96.01 & 0.17 & 0.01 & 0.00 & 1.32 & 0.11 & 2.38 & Disclosure & 96.01 \\ 
            Topic 16 & 6.39 & 0.39 & 0.66 & 29.80 & 25.70 & 29.26 & 7.82 & Market Update & 29.80 \\ 
            Topic 17 & 98.86 & 0.00 & 0.00 & 0.03 & 0.15 & 0.05 & 0.91 & Disclosure & 98.86 \\ 
            Topic 18 & 98.67 & 0.00 & 0.02 & 0.16 & 0.10 & 0.53 & 0.52 & Disclosure & 98.67 \\ 
            Topic 19 & 63.62 & 10.74 & 0.00 & 0.05 & 22.38 & 1.73 & 1.49 & Disclosure & 63.62 \\ 
            Topic 20 & 99.99 & 0.00 & 0.00 & 0.01 & 0.00 & 0.00 & 0.00 & Disclosure & 99.99 \\ 
            \hline
            \end{tabular}
            \caption{Classification accuracy and content distribution for topics identified by the Top2Vec 20 model applied to hedge fund text. This table presents the percentage breakdown of text samples within each topic across six predefined document categories, plus an "other" category for content that doesn't fit these classifications. Each row represents one of the 20 topics extracted by LDA 20, while columns represent the document categories: "Disclosure" (legal disclaimers and regulatory statements), "Fund Terms" (fee structures and investment conditions), "Investment Team" (personnel information), "Market Update" (economic and market commentary), "Performance Commentary" (fund results and attribution), and "Strategy Overview" (investment approach and philosophy). The "Predicted Class" column indicates the dominant category for each topic based on the highest percentage value in the row, while the "Accuracy" column shows the corresponding percentage of correctly classified text samples within that topic.}
            \caption{metrics of top2vec 20}
        \label{tab:metrics_t2v_20}
    \end{table}
    \end{landscape}

        \begin{landscape}
    \begin{table}
    \scriptsize
        \centering
        \begin{tabular}{lcrcc}
                \hline

            No. & Top-10 keywords & \#paragraphs & LLM \& Manual Annotation\\
                    \hline

            Topic 1 & \makecell{investmen, investcorp, financier, financiers, investors, investor, theinvestment, financieres, securities, analysts} & 13,408 &  Investment Team\\
            
            Topic 2 & \makecell{investcorp, securities, investor, investorrights, ltda, theinvestment, investors, ltd, investm, financials} & 9,238 &  Disclosure\\
            
            Topic 3 & \makecell{investability, investment, investments, invest, investible, investor, investable, securities, invests, investm} & 8,859 &  Disclosure\\
            
            Topic 4 & \makecell{securities, investor, investorrights, financials, financial, theinvestment, investments, investors, invest, ltda} & 8,677 &  Disclosure\\
            
            Topic 5 & \makecell{investment, investments, portfolios, funds, fund, investability, portfolio, financials, invests, underinvestment} & 8,600 &  Disclosure\\
            
            Topic 6 & \makecell{fees, expenses, expense, costing, underinvestment, funds, financings, fund, expenditure, theinvestment} & 8,518 &  Disclosure\\
            
            Topic 7 & \makecell{investor, investorrights, aifs, aifmd, investees, investee, investors, investm, etf, aifm} & 8,287 &  Disclosure \\
            
            Topic 8 & \makecell{disclosures, disclosure, disclaimers, disclose, disclaims, validity, information, warranties,\\ confidentiality, uncertainties} & 8,281 &  Disclosure\\
            
            Topic 9 & \makecell{securities, investor, investorrights, invest, investments, theinvestment, invests, investors, investment, investible} & 8,150 &  Disclosure\\
            
            Topic 10 & \makecell{securities, investorrights, investor, theinvestment, noninvestment, investors, nvestments, nvestment, invest} & 8,142 &  Disclosure\\
            
            Topic 11 & \makecell{financings, finance, theinvestment, financials, financing, reinvestments, underinvestment, recessions,\\ subinvestment, financial} & 7,654 &  Market Update\\

            Topic 12 & \makecell{investcorp, reinvestments, theinvestment, reinvestment, investors, underinvestment, market, stock,\\ investability, prospects} & 7,377 &  Performance Commentary \\

            Topic 13 & \makecell{investability, investments, investment, invests, investible, theinvestment, invested, securities,\\ invest, investmentgrade} & 7,259 &  Disclosure\\

            Topic 14 & \makecell{investments, invest, investment, investability, invests, investing, investable, investors, investible, investor} & 7,247 & Disclosure\\

            Topic 15 & \makecell{investments, investors, investability, investment, investor, investing, invest, invests, assetmanagement} & 7,153 & Strategy Overview\\

            Topic 16 & \makecell{earnings, quarterly, financials, investmentgrade, valuations, annum, nvestment, investment, nvestments} & 6,826 & Disclosure\\

            Topic 17 & \makecell{invest, investor, investment, investible, invested, invests, investable, investments, investorrights, investm} & 6,717 & Disclosure\\

            Topic 18 & \makecell{investments, theinvestment, investors, reinvestments, invest, investment, invests, reinvestment, investing, investor} & 6,693 & Performance Commentary\\

            Topic 19 & \makecell{disclosures, disclosure, disclose, securities, investorrights, theinvestment, confidentiality, investor, noninvestment} & 6,670 & Disclosure\\

            Topic 20 & \makecell{investments, investors, invest, investment, invests, investor, securities, portfolios, fund, theinvestment} & 6,641 & Strategy Overview\\

                    \hline

        \end{tabular}
        \caption{Sample topics generated by Top2Vec after reducing the number of topics to 100. The table presents the top words for each topic, the number of assigned paragraph text, and representative annotations.}
        \label{tab:topics_top2vec_100}
    \end{table}
    \end{landscape}

        \begin{landscape}
    \begin{table}
    \tiny
        \centering
            \begin{tabular}{lccccccclc}
            \hline
            No. & Disclosure & Fund Terms & Investment Team & Market Update & Performance Commentary & Strategy Overview & other & Predicted Class & Accuracy (\%) \\ \hline
            Topic 1 & 0.45 & 0.00 & 97.05 & 0.06 & 0.33 & 0.72 & 1.40 & Investment Team & 97.05 \\ 
            Topic 2 & 98.69 & 0.05 & 0.08 & 0.05 & 0.03 & 0.56 & 0.54 & Disclosure & 98.69 \\ 
            Topic 3 & 99.83 & 0.00 & 0.00 & 0.00 & 0.00 & 0.10 & 0.07 & Disclosure & 99.83 \\ 
            Topic 4 & 99.91 & 0.04 & 0.00 & 0.02 & 0.04 & 0.00 & 0.00 & Disclosure & 99.91 \\ 
            Topic 5 & 85.41 & 2.01 & 0.00 & 0.00 & 12.39 & 0.18 & 0.00 & Disclosure & 85.41 \\ 
            Topic 6 & 69.06 & 25.87 & 0.00 & 0.06 & 4.63 & 0.06 & 0.32 & Disclosure & 69.06 \\ 
            Topic 7 & 100.00 & 0.00 & 0.00 & 0.00 & 0.00 & 0.00 & 0.00 & Disclosure & 100.00 \\ 
            Topic 8 & 99.71 & 0.00 & 0.00 & 0.00 & 0.00 & 0.00 & 0.29 & Disclosure & 99.71 \\ 
            Topic 9 & 99.91 & 0.09 & 0.00 & 0.00 & 0.00 & 0.00 & 0.00 & Disclosure & 99.91 \\ 
            Topic 10 & 99.74 & 0.22 & 0.00 & 0.00 & 0.00 & 0.00 & 0.04 & Disclosure & 99.74 \\ 
            Topic 11 & 1.36 & 0.00 & 0.00 & 65.91 & 20.34 & 11.48 & 0.91 & Market Update & 65.91 \\ 
            Topic 12 & 0.74 & 0.00 & 1.03 & 11.65 & 58.55 & 17.70 & 10.32 & Performance Commentary & 58.55 \\ 
            Topic 13 & 98.36 & 0.00 & 0.00 & 0.00 & 1.54 & 0.00 & 0.10 & Disclosure & 98.36 \\ 
            Topic 14 & 99.73 & 0.19 & 0.00 & 0.00 & 0.00 & 0.00 & 0.08 & Disclosure & 99.73 \\ 
            Topic 15 & 13.87 & 0.07 & 23.87 & 0.21 & 2.11 & 42.96 & 16.90 & Strategy Overview & 42.96 \\ 
            Topic 16 & 74.52 & 1.62 & 0.00 & 0.00 & 17.67 & 0.81 & 5.38 & Disclosure & 74.52 \\ 
            Topic 17 & 99.89 & 0.05 & 0.00 & 0.00 & 0.00 & 0.00 & 0.05 & Disclosure & 99.89 \\ 
            Topic 18 & 8.84 & 7.36 & 0.29 & 3.98 & 55.67 & 19.73 & 4.12 & Performance Commentary & 55.67 \\ 
            Topic 19 & 94.60 & 5.22 & 0.00 & 0.00 & 0.00 & 0.14 & 0.05 & Disclosure & 94.60 \\ 
            Topic 20 & 12.14 & 9.29 & 4.02 & 0.19 & 3.78 & 54.12 & 16.47 & Strategy Overview & 54.12 \\ 
            \hline
            \end{tabular}
            \caption{Classification accuracy and content distribution for topics identified by the Top2Vec 100 model applied to hedge fund text. This table presents the percentage breakdown of text samples within each topic across six predefined document categories, plus an "other" category for content that doesn't fit these classifications. Each row represents one of the 20 topics extracted by LDA 20, while columns represent the document categories: "Disclosure" (legal disclaimers and regulatory statements), "Fund Terms" (fee structures and investment conditions), "Investment Team" (personnel information), "Market Update" (economic and market commentary), "Performance Commentary" (fund results and attribution), and "Strategy Overview" (investment approach and philosophy). The "Predicted Class" column indicates the dominant category for each topic based on the highest percentage value in the row, while the "Accuracy" column shows the corresponding percentage of correctly classified text samples within that topic.}
            \caption{metrics of top2vec 100}
        \label{tab:metrics_t2v_100}
    \end{table}
    \end{landscape}

\section{Declaration of generative AI and AI-assisted technologies in the writing process}
During the preparation of this work, the author used ChatGPT in order to improve language and readability. After using this tool, the author reviewed and edited the content as needed and takes full responsibility for the content of the published article.  

\bibliographystyle{elsarticle-harv} 
	\bibliography{ref.bib}
\end{document}